%% file: template.tex
\documentclass[11pt]{article}

\usepackage{mathtools}
\usepackage{amssymb}
\usepackage{amsmath}
\usepackage{amsthm}
\usepackage{cancel}
\usepackage{tikz}
\usepackage{arydshln}
\usepackage{hyperref}
\usepackage[margin=3cm]{geometry}

\newcommand\matlab{\textsc{Matlab}}
\newcommand\numpy{\textsc{NumPy}}

\newcommand\gitlab{\textsc{GitLab}}

\newcommand{\R}{\mathbb{R}}
\newcommand{\N}{\mathbb{N}}
\newcommand{\E}{\mathbb{E}}
\newcommand{\V}{\mathbb{V}}

\makeatletter \renewcommand\d[1]{%
  \mathop{}\!\mathrm{d}#1}
\makeatother
\newcommand{\argmin}{\operatornamewithlimits{arg\;min}}

\title{Efficient Uncertainty Quantification and Sensitivity Analysis in Epidemic Modelling using Polynomial Chaos}

\author{Bjørn Jensen$^{\rm{a}}$\footnote{bjorn.jensen@helsinki.fi; \url{https://orcid.org/0000-0002-4743-2631}}, Allan P. Engsig-Karup$^{\rm{b}}$\footnote{apek@dtu.dk; \url{https://orcid.org/0000-0001-8626-1575}}\ \  and Kim Knudsen$^{\rm{b}}$\footnote{kiknu@dtu.dk; \url{https://orcid.org/0000-0002-4875-3074}}
	\\ 
	\\ $^{\rm{a}}$Department of Mathematics and Statistics\\
	University of Helsinki\\
	00560 Helsinki, Finland
	\\ 
	\\ $^{\rm{b}}$Department of Applied Mathematics and Computer Science\\
	Technical University of Denmark\\
	2800 Kgs.\ Lyngby, Denmark}

\begin{document}

\input{doc_content}

\end{document}

%% file: doc_content.tex
\maketitle

\begin{abstract}
In the political decision process and control of COVID-19 (and other epidemic diseases), mathematical models play an important role. It is crucial to understand and quantify the uncertainty in models and their predictions in order to take the right decisions and trustfully communicate results and limitations. We propose to do uncertainty quantification in SIR-type models using the efficient framework of generalized Polynomial Chaos. Through two particular case studies based on Danish data for the spread of Covid-19  we demonstrate the applicability of the technique. The test cases are related to peak time estimation and superspeading and illustrate how very few model evaluations can provide insightful statistics.

\end{abstract}

\section{Introduction}
Quantification of uncertainty is an important aspect in all model and data driven problems. When a computed solution relies on the collection of imperfect data, the result is rarely perfect; the solution rather represents an estimate of some desired value. Also the model used on the problem may not represent the full phenomenon, either from deliberate simplifications or due to complicated mechanisms beyond our current understanding. Sources of uncertainty are commonly classified as being either aleatoric or epistemic uncertainty; the former classify uncertainties impossible to know due to insufficient understanding or perhaps measurement errors at a currently unreachable scale, and the latter encapsulates for example the deliberate reductions in precision due to simplified models or for instance less data collection. In either case, the uncertainty will influence the credibility of the solution of the problem and quantifying that uncertainty helps ascertain the trust we should have, or the risk we take, when making decisions based on such models. Thus it plays a key role in both problems about prediction and simulation of potential scenarios. The use of computational methods in this study is commonly referred to as uncertainty quantification (UQ).

In this manuscript we demonstrate how techniques from uncertainty quantification apply to epidemic modelling to provide insight and locate the key uncertainties; i.e. which data sources provide the biggest uncertainties. Such knowledge is crucial for mitigation strategies, restriction policies, etc. targeting controlling or reducing the impact of the spread of diseases for securing public health. This will in part also improve the ability to deal with uncertainty in predictive modelling.

Uncertainty quantification as an independent field grew out of problems in various other fields such as probability theory, dynamical systems and numerical simulations. Sampling based techniques, such as Markov Chain Monte Carlo (MCMC) methods and bootstrapping, have seen use in epidemic modelling as seen in the studies~\cite{sneppen2020impact,TAGHIZADEH2020104011,house2016bayesian}, and by the expert group\footnote{https://covid19.ssi.dk/analyser-og-prognoser/modelberegninger (accessed April 9th, 2021; in Danish)} providing the Covid-19 related modelling for the Danish government.  We propose an alternative approach called generalized Polynomial Chaos~\cite{ghanem1990polynomial,xiu2002wiener,xiu2010numerical,bigoni2015uncertainty} as an efficient general non-iterative framework to do UQ-analysis using forward modelling where the uncertainties are parameterized; the outcome being a prediction in terms of the solution's expected value and uncertainty in terms of the solution's variance.

In epidemic modelling the spreading of an infectious disease is investigated through the application of mathematical models. Models of various complexity, flexibility, restrictions and assumptions exist. If appropriately combined with data the models, within their assumptions, provide insight into the behaviour of the disease within the population. It may yield estimates for its duration, the peak infection and various other aspects. In this paper we use extended versions of the SIR-type model, which is the most common epidemic model.  

As mentioned, such models only work within their assumptions and are thus not perfect descriptors. They depend on a limited set of parameters, which have to be calibrated matching the model to the available data. Practical data sources come with their own randomness and incompleteness, so it is typical to attempt to capture the broader trends rather than say day-to-day fluctuations. This typically manifests as smoothness in the models and retaining a coarse state space. Parameter fitting under these constraints may lead to some level of confidence in the parameters, which can be mathematically represented by a probability distribution.

\begin{figure}[!ht]
\begin{center}
    \includegraphics[]{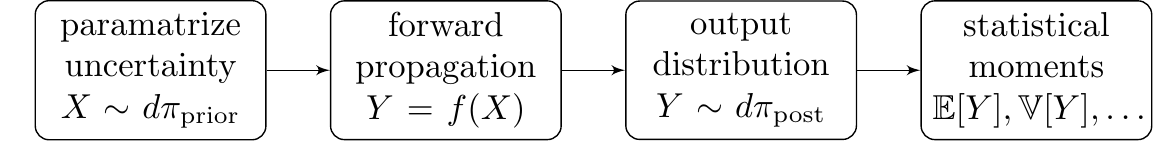}
\end{center}
\caption{The workflow used for handling UQ.}
\label{fig:uq-flow}
\end{figure}

With a probability distribution, in place the uncertainty can be propagated to the model output using UQ techniques providing a distribution on the output as illustrated in Figure \ref{fig:uq-flow}. This allows for computation of various statistics on the output, e.g. mean, variance, confidence intervals, etc. This is not trivial to do, however, since quantifying the uncertainty in a prediction comes at a cost.

Common techniques are MCMC methods, which rely on sampling for exploring the potentially complicated probability distribution of the prediction. However, sampling requires model evaluations, which can be expensive as a vast number of samples are necessary for MCMC due to its slow convergence rate. 

Generalized Polynomial Chaos (gPC) poses an efficient alternative non-sampling based method, which can provide very good estimates using significantly fewer model evaluations when the dimension of the problem is sufficiently low. Provided that the number of uncertain parameters is sufficiently low, it is a very efficient method. The drawback is that it suffers from the curse of dimensionality, when the parameter count (i.e. the dimension) grows, and requires smoothness of the prediction distribution. It utilizes orthogonal polynomials and Gaussian quadrature to optimize the number of model evaluations necessary to compute statistics by means of an orthonormal expansion. gPC has also been used on Spanish data in \cite{olivares2021uncertainty}.

Once computation of various statistics given uncertainty in inputs are in place, we obtain information about the stochasticity of our results. An interesting question is then where we would gain the most from building confidence in an input parameter? In other words, are some parameters significantly more contributing to the uncertainty in the model output? Sobol indices provide an insight in this regard. This is called Variance-based sensitivity analysis. Sobol indices have for example been applied to the Bristish \textsc{CovidSim} model in \cite{edeling2021impact}.

This manuscript is structured as follows: Section 2 and 3 provides the theoretical background. In Section 2 we give an introduction to Polynomial Chaos and illustrate how various basic statistics are directly computable from the expansion coefficients. Sobol indices are given a brief introduction in Section 3. We provide a short derivation of their formulation and relate them back to the Polynomial Chaos by providing formula for their computation in terms of the expansion coefficients. These sections are based on the expositions in \cite{sudret2008global,alexanderian2020variance}.

The main novelty of our work is in Section 4, where we demonstrate the utility of gPC analysis and Sobol indices in epidemic models and apply them to Danish data from the early phases of Coronavirus SARS-CoV-2 (Covid-19). The versatility of the tools is illustrated in two different cases. Case 1 is a simple SIR-model based estimation of the timing and size of the peak of an epidemic. Case 2 attempts to provide a way of modelling superspreaders in SIR-type models inspired by the recent manuscript \cite{sneppen2020impact}.

The computations included in this manuscript were done in \matlab{} and the framework is available as a small toolbox on the DTU  \gitlab{} server \footnote{\url{https://gitlab.gbar.dtu.dk/bcsj/covid-19-ctrl-public-code}}. The methods used for computing the various quadratures have been ported to \matlab{} from \numpy{}\cite{oliphant2006guide}.

\section{Polynomial Chaos Expansion}
We will consider a model described by the input-output map $ f\colon \Omega \subseteq \mathcal X \to \mathcal Y $, where $\mathcal X=\R^n $ is the parameter space and $ \Omega $ is a subset and $ \mathcal Y=\R^m $ is the output space. The aim is to quantify behaviour in the model $ Y = f(X) \in \mathcal Y $ under some variation of the parameter $ X \in \mathcal X $. There are various ways of approaching this. We could compute derivatives $ \mathrm{d}f\colon \mathcal X \to \mathcal L(\mathcal X,\mathcal Y) $, $ \mathrm{d}^2\!f $, $\dots$, $ \mathrm{d}^k\!f $, etc. However, unless $ f $ is linear this information is exclusively local in nature and does not explain global trends. Further more, we will consider $ f $ as a black box in general so we cannot assume any directly exploitable structure.

A way to capture more broad information about $ f $ is to consider its coefficients with respect to a suitable basis. A good choice of basis yields a fast decay in the coefficients of $ f $, which leads to a good approximation by a finite series representation. Of course, no basis will display a fast coefficient decay for every decomposable function, however, there are choices applicable for fairly broad and useful classes of functions.

Polynomial Chaos decomposes $ f \in L^2(\Omega, \mathcal Y, \d\mu) $ in a basis of orthonormal polynomials $ \{\phi_\alpha\} $ of increasing order, where $ \alpha = (\alpha_1,\dots,\alpha_n) \in \N^n $ is a multi-index. We will use the notational convention that $ \alpha = 0 $ when $ \alpha_i = 0 $ for all $ 1\leq i\leq n $. Note that $ \phi_0(x) = 1 $; the zeroth order polynomial is always constant. The decomposition is standard
\begin{equation}
    f(X) = \sum_{0\leq\alpha} \langle f,\phi_\alpha\rangle_\mu \phi_\alpha(X), \quad \text{where} \quad \langle f,\phi_\alpha\rangle_\mu = \int_\Omega f(x)\phi_\alpha(x)\d\mu(x).
\end{equation}
Here $ \langle f,\phi_\alpha\rangle_\mu $ are the coefficients of $ f $. In practice the sum is truncated and as mentioned above approximated by a finite series representation. This is reasonable since for $ \{\phi_\alpha\} $ orthonormal in $ L^2(\Omega,\mathcal Y, \d\mu) $ the coefficients decays towards zero, and assuming $ f $ is well-behaved this decay is fast.

For a number of common probability measures $ \d\mu $ the orthonormal polynomials are well-known and easy to generate. They also yield Gaussian quadratures with respect to these probability measures, which makes the computation of the involved integrals fast.

Consider for instance the standard normal distribution $ \mathcal N(0,1^2) $, which up to a scaling constant has probability measure $ \exp(-x^2/2) $. The (probabilists) Hermite polynomials form an orthogonal sequence with respect to this measure. Picking a degree $ n_\text{quad} $ and computing the roots $ \xi_i $ of the Hermite polynomial of the corresponding the degree together with the weights $ w_i $ gives a quadrature rule for integration
\[ \int_\R f(x)e^{-\frac{x^2}{2}}\d x \approx \sum_{i=1}^{\mathclap{n_\text{quad}}}w_i f(\xi_i), \]
where the equality is exact whenever $ f $ is a polynomial with $ \deg(f) \leq 2n_{\text{quad}}-1 $.

\subsection{Statistical properties}
While stochastic phenomenons come with expressive distributions, which are detailed, we will often quantify an uncertain output $ Y $ in terms of the basic statistical properties like the mean, variance and covariance, as these are easier to process. Given a model $ f $ with parameters characterized by the random variable $ X $, the resulting output $ Y = f(X) $ is a new random variable and its basic statistical properties become directly computable from the coefficients in our polynomial expansion for $ f $.

Assume that $ \d\mu $ is a probability measure on the parameter set $ \Omega $, and that $ \{\phi_\alpha\} $ is an orthonormal basis as above.
Consider the random variable $ Y = f(X) $, where $ X \sim \d\mu $, i.e. it follows the distribution defined by $ \d\mu $. Let us denote by $ c_\alpha = \langle f,\phi_\alpha\rangle_\mu $, then it is easy to see that we immediately obtain the mean value in terms of the first coefficient;
\begin{equation}
    \E[Y] = \int_\Omega f(x) \d\mu(x) = \int_\Omega f(x) \phi_0(x)\d\mu(x) = c_0,
\end{equation}
and we can do similarly for other statistics.

The variance may be derived as
\begin{equation}
    \V[Y] = \sum_{0\leq\alpha} c_\alpha^2\V\left[\phi_\alpha(X)\right] = \sum_{0<\alpha} c_\alpha^2,
\end{equation}
using $ \E[\phi_\alpha] = \delta_\alpha $ and $ \E[\phi_\alpha\phi_\beta] = \delta_{\alpha-\beta} $ by orthonormality.

Consider now the random variables $ Y_1 = f_1(X) $ and $ Y_2 = f_2(X) $ with coefficients $ \{c_\alpha\} $ and $ \{d_\alpha\} $, then a computation analogous to that for the variance yields the covariance as
\begin{equation}
    \operatorname{Cov}(Y_1,Y_2) = \E[Y_1Y_2] - \E[Y_1]\E[Y_2] = \sum_{0<\alpha}^\infty c_\alpha d_\alpha.
\end{equation}
In fact, if $ \mathbf Y $ is a random vector with $ Y_i = f_i(X) $, $ 1 \leq i \leq k $ and we have coefficients $ c_{i,\alpha} = \langle f_i,\phi_\alpha\rangle_\mu $. Forming the infinite matrix
\[ 
    Q = \begin{bmatrix}
    c_{1,\alpha(1)} & c_{1,\alpha(2)} & \cdots & c_{1,\alpha(j)} & \cdots\\
    c_{2,\alpha(1)} & c_{2,\alpha(2)} & \cdots & c_{2,\alpha(j)} & \cdots\\
    \vdots & \vdots & & \vdots & \\
    c_{k,\alpha(1)} & c_{k,\alpha(2)} & \cdots & c_{k,\alpha(j)} & \cdots
\end{bmatrix} \in \R^{k\times\N}, 
\]
where $ \alpha(\cdot)\colon \N \to \N^n $ is some traversal of the multi-index space with $ \alpha(0) = (0,0,\dots,0) $, the covariance matrix $ C = \operatorname{Cov(\mathbf Y,\mathbf Y)} $ is of the form $ C = QQ^T \in \R^{k\times k} $.

\section{Sobol indices}
As we often quantify our uncertain output $ Y = f(X) $ in terms of the statistical properties, it is natural to ask which of the components among the parameters $ X $ produced the largest contribution to the variance in quantities of interest. In other words, if we may somehow reduce the uncertainty in a single parameter, which choice would yield the greatest decrease in the uncertainty in the output? It is important to keep in mind that even if a single parameter carries a huge uncertainty it might not be very influential in the model. This kind of insight may be utilized to save on computational effort and to identify the most influential parameters.

The Sobol indices form a quantification of the variance contribution on the output $ Y $ from each individual parameter and each combination of parameters in $ X $. Like the basic statistical properties, the Sobol indices are computable from the polynomial expansion coefficients for $ f $. We give a brief example here, then present the formulation of the Sobol indices, and follow up with the derivation in terms of the coefficients.

Consider the map $ f\colon \Omega \subset \mathcal X \to \mathcal Y $, $ \mathcal X =\R^n $. Though we could in principle consider $ \mathcal Y $ as a vector space, due to the rapid growth in required number of indices complicating the notation we shall refrain and instead have $ \mathcal Y = \R $.

Let $ \d\mu = \prod_{i=1}^n\d\mu_i $ be a probability measure on $ \Omega\subseteq \mathcal X $ and $ X = (X_1,\dots,X_n) $, $ X_i \sim \d\mu_i $. In essence the Sobol indices is a decomposition of the variance of the output $ Y = f(X) $ in terms of the different combinations of input parameters $ X_1,\dots,X_n $. 

We consider a few simple scenarios. Let $ X_1 \sim \mathcal N(0,a^2) $ and $ X_2 \sim \mathcal N(0,b^2) $ be normally distributed. Say $ Y = X_1 + X_2 $, then the Sobol indices would be $ S_1, S_2 $ and $ S_{12} $ corresponding to each non-empty combination of $ X_1 $ and $ X_2 $. Their values would be $ S_1 = \frac{a^2}{a^2+b^2} $, $ S_2 = \frac{b^2}{a^2+b^2} $ and $ S_{12} = 0 $. In other words, say $ a > b $, then it is better to decrease the uncertainty in $ X_1 $ rather than $ X_2 $.

In contrast, consider $ Y = X_1X_2 $, then the Sobol indices are $ S_1 = S_2 = 0 $ and $ S_{12} = 1 $, so it decreasing the uncertainty in either is equally beneficial.

\subsection{Formulation}
To compute the Sobol indices we rely on a decomposition into marginalizations of $ f $. We give a derivation of the Sobol indices based on the exposition in~\cite{sudret2008global} to make clear their computation.

Let $ U = \{1,\dots,n\} $.
\begin{equation}
    f(X) = \sum_{\mathclap{u\subseteq U}} f_u(X_u), \label{eq:sobol-exp-f}
\end{equation}
where $ X_u = (X_i)_{i\in u} $ and $ f_\emptyset(X_\emptyset) \coloneqq f_0 = \E[f(X)] $. The remaining functions $ f_u $, $ u \neq \emptyset $, are then recursively defined by
\begin{equation}
    f_u(X_u) = \E_{U\backslash u}[f(X)] - \sum_{u'\subsetneq u} f_{u'}(X_{u'}), \label{eq:sobol-exp-recurs}
\end{equation}
where $ \E_{U\backslash u}[f(X)] $ is a marginalization, i.e.
\begin{equation}
    \E_u[f(X)] \coloneqq \int_{\R^k} f(x)\prod_{i\in u}\d\mu_i(x_i), \quad k = |u| \label{eq:sobol-exp-margin}
\end{equation}
with $ \d\mu = 0 $ outside $ \Omega $.

Note that the sum is telescopic, each component corresponding to a set $ u $ subtracting subset components again. Hence we may compute each $ f_u(X_u) $ starting from the smallest subsets of $ u $ and progressively building up to the bigger subsets.

We consider now the variance of $ f(X) $ and apply the expansion \eqref{eq:sobol-exp-f} to obtain
\begin{equation}
    \V[f(X)] = \sum_{u\subseteq U, u\neq\emptyset}\V[f_u(X_u)]. \label{eq:sobol-exp-var}
\end{equation}
By dividing by the left hand side in \eqref{eq:sobol-exp-var} we get 
\begin{equation}
    1 = \sum_{u\subseteq U, u\neq\emptyset}\frac{\V[f_u(X_u)]}{\V[f(X)]} = \sum_{u\subseteq U,u\neq\emptyset} S_u,
\end{equation}
defining $ S_u \coloneqq \V[f_u(X_u)]/\V[f(X)] $. The Sobol indices are then $ \{S_u\}_{u\subseteq U,u\neq\emptyset} $.

\subsection{Relation to Polynomial Chaos}
The Sobol indices are efficiently computable from the PC coefficients. The marginalizations of the distribution arise as restrictions to certain subsets of the coefficients.

Let $ c_\alpha $ be the PC coefficients of $ f $. To compute the Sobol indices we wish to compute the terms $ \V[f_u(X_u)] $. Taking the variance on both sides in \eqref{eq:sobol-exp-recurs} we get
\[
    \V[f_u(X_u)] = \V\left[\E_{U\backslash u}[f(X)]\right] - \sum_{u'\subsetneq u} \V[f_{u'}(X_{u'})].
\] 
Clearly, if we compute bottom up hierarchically using the partial ordering $ u \leq v $ if $ u \subseteq v $, we simply need to compute the marginalizations $ \V[\E_{U\backslash u}[f(X)]] $ and then subtract formerly computed values. 

Due to the maginalizations of $ f $ we will need to consider the marginal structure of our basis functions $ \{\phi_\alpha\} $ too. For a multi-index $ \alpha \in \N^n $ we shall use the notation
\[
    \phi_\alpha(x) = \psi_{1,\alpha_1}(x_1)\cdots \psi_{n,\alpha_n}(x_n),
\]
where $ \{\psi_{i,j}\}_j $ is the orthonormal polynomial basis for parameter $ X_i $. With this we may derive
\begin{align*} 
    \E_{U\backslash u}[f(X)] &= \int_{\R^k} f(x)\prod_{i\in u}\d\mu_i(x_i) \\
    &= \int_{\R^k} \sum_{0\leq \alpha}c_\alpha \phi_\alpha(x)\prod_{i\in u}\d\mu_i(x_i) \\
    &= \sum_{0\leq \alpha}c_\alpha \left(\prod_{i\in U\backslash u}\psi_{i,\alpha_i}(X_i)\right)\left(\prod_{i\in u}\int_{\R}\psi_{i,\alpha_i}(x_i)\d\mu_i(x_i)\right) \\
    &= \sum_{0\leq \alpha}c_\alpha \left(\prod_{i\in U\backslash u}\psi_{i,\alpha_i}(X_i)\right)\left(\prod_{i\in u}\E[\psi_{i,\alpha_i}(X_i)]\right)\\
\intertext{(note that this product of mean values is 0 unless $ \alpha_i = 0 $ for all $ i \in u $; we write simply $ \alpha_u = 0 $)}
    &= \sum_{0\leq \alpha, \alpha_u=0}c_\alpha \prod_{i\in U\backslash u}\psi_{i,\alpha_i}(X_i) \\
\intertext{(as $ \psi_{i,0}(x)=1$ this product extends to all of $ \alpha $ again now that $ \alpha_u = 0 $ is fixed)}
    &= \sum_{0\leq \alpha, \alpha_u=0}c_\alpha^2 \phi_{\alpha}(X).
\end{align*}
Taking the variance of the above and using the fact that $ \V[\phi_\alpha(X)] = 1 $ for $ \alpha \neq 0 $ and zero otherwise we get
\begin{equation}
    \V\left[\E_{U\backslash u}[f(X)]\right] = \sum_{0\leq \alpha, \alpha_u=0}c_\alpha^2 \V[\phi_{\alpha}(X)] = \sum_{0<\alpha, \alpha_u=0}c_\alpha^2
\end{equation}

Visually, if we consider just two parameters, we see in the coefficient grid below how the different coefficients distribute themselves among the 
\begin{equation*}
    \begingroup
    \setlength\arraycolsep{7pt}
    \renewcommand{\arraystretch}{2}
    \begin{array}{cc:ccccc}
        & \cancel{c_{0,0}^2} & c_{0,1}^2 & c_{0,2}^2 & c_{0,3}^2 & \cdots & \sum \square = \V[f_2(x_2)]\\ \hdashline
        &c_{1,0}^2 & c_{1,1}^2 & c_{1,2}^2 & c_{1,3}^2 & \cdots &\\
        &c_{2,0}^2 & c_{2,1}^2 & c_{2,2}^2 & c_{2,3}^2 & \cdots &\\
        &c_{3,0}^2 & c_{3,1}^2 & c_{3,2}^2 & c_{3,3}^2 & \cdots &\\
        &\vdots & \vdots & \vdots & \vdots & \ddots & \\
        & \mathclap{\text{$\sum \square= \V[f_1(x_1)]$\phantom{abcdefghijklm}}} & & & & & \text{\rotatebox{0}{$\sum \square= \V[f_{12}(x_{12})]$}}
    \end{array}
    \endgroup
\end{equation*}
Here ``$ \sum\square $'' is simply intended as a placeholder symbol for the \emph{sum of each of the elements in the box}. Note that $ c_{0,0} $ is crossed out, as it is the mean, which does not contribute the the variance.

\section{Uncertainty Quantification in modelling spread of diseases using Polynomimal Chaos}
In this section we present two cases to demonstrate the flexibility of the above techniques by applying them to SIR-type models. The first case considers a SEIR-model and compute distributions for the size and timing of the peak of the modelled epidemic.

For the second case a more extensive SIR-type model is considered. Inspired by the agent based modelling of superspreaders discussed in \cite{sneppen2020impact} we construct a multi-compartment SIR-type model and formulate a modelling approach for superspreaders leading to comparable results despite the differences in modelling assumptions. With this model we perform a UQ analysis on the coefficients modelling the government imposed restrictions.

In both cases the population size $ N $ is taken as $ 5.8\times 10^6 $ matching the size of the Danish population.

\subsection{Case 1: Epidemic peak}
\begin{figure}[!ht]
\begin{center}
    \begin{tikzpicture}[scale=1,transform shape]
        \newcommand\s{1.75}
        \node[shape=circle,draw=black,minimum size=1cm,thick] (S) at (0,0) {$ S $};
        \node[shape=circle,draw=black,minimum size=1cm,thick] (E) at (\s,0) {$ E $};
        \node[shape=circle,draw=black,minimum size=1cm,thick] (I) at (2*\s,0) {$ I $};
        \node[shape=circle,draw=black,minimum size=1cm,thick] (R) at (3*\s,0) {$ R $};
        
        \path [->,thick, draw=black] (S) edge node[anchor=south,yshift=.3cm] {$\beta$} (E);
        \path [->,thick, draw=black] (E) edge node[anchor=south,yshift=.3cm] {$\sigma$} (I);
        \path [->,thick, draw=black] (I) edge node[anchor=south,yshift=.3cm] {$\gamma$} (R);

        \path [->,thick, draw=black, bend left=65, dashed] (I) edge (.5*\s,-.15*\s);
        
    \end{tikzpicture}
\end{center}
\caption{Illustration of compartments and transmission rates for a SEIR model.}
\label{fig:seir-model-compartments}
\end{figure}
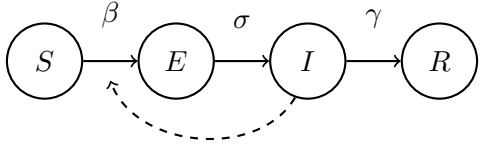

For this simple case we consider an SEIR-model, i.e. a model with the compartments (S)usceptible, (E)xposed, (I)infectious and (R)ecovered/removed. The model is visualized in the diagram in Figure \ref{fig:seir-model-compartments}.

As an ODE system the model is of the form
\begin{subequations}
    \begin{align}
        \frac{\partial S}{\partial t} &= -\beta \frac{I(t)S(t)}{N}, \label{eq:seir-ds}\\
        \frac{\partial E}{\partial t} &= \beta \frac{I(t)S(t)}{N} - \sigma E(t), \\
        \frac{\partial I}{\partial t} &= \sigma E(t) - \gamma I(t), \\
        \frac{\partial R}{\partial t} &= \gamma I(t),
    \end{align}    
\end{subequations}
where $ \beta $, $ \sigma $ and $ \gamma $ are transition coefficients and $ N $ the total size of the population. $ \sigma $ is the rate at which people progress from being exposed (incubating) to becoming infectious individuals, and $ \gamma $ is the rate at which one recovers (or dies) from the disease. Their reciprocals are the average time an individual spends in the exposed and infectious compartments respectively. $ \beta $ denotes the average rate of infection happening in the population. This quantity is a function of the infectiousness of the virus and the social patterns of the population; e.g. higher hygiene standard in the population would lead to a lower $ \beta $. Note that it is assumed that $ S+E+I+R = N $ at all times, which is typically used for shorter time horizons in the modelling.

The model makes the assumptions that we are dealing with a large population with heterogeneous mixing; in other words any randomly sampled subset of individuals from the population should behave the same at the  macroscopic level of a society.

The progress of an epidemic can roughly be modelled this way. The model is easy to expand in complexity to incorporate various sources of data and phenomenons. We see this in the following case. 

In an epidemic the number of infected individuals will rise rapidly as each infected individual will infect several others. However, as the population becomes saturated with infected individuals the likelihood of a meeting between an infected and a susceptible will decrease. We say \emph{herd immunity} is kicking in. Hence, the epidemic peaks at some time $ t_{\text{peak}} $ where the number of infectious individuals are at its highest.

We consider in this example each parameter $ \beta $, $ \sigma $ and $ \gamma $ uncertain. The uncertainties are given as uncertainty in the reproduction number $ R_0 = \frac{\beta}{\gamma} $, in the duration in the exposed compartment $ \tau_{\text{inc}} = \sigma^{-1} $ and the duration in the infectious compartment $ \tau_{\text{inf}} = \gamma^{-1} $. As these are positive quantities we assume each log-normally distributed. We thus consider the map
\begin{equation}
    \mathcal F \colon (R_0,\tau_{\text{inc}},\tau_{\text{inf}}) \mapsto (t_{\text{peak}}, I_{\text{peak}}),
\end{equation}
where $ I_{\text{peak}} \coloneqq I(t_{\text{peak}}) $. As the log-normal distribution is simply a transformation of the normal distribution, it is a simple task to transform the quadrature nodes accordingly.

We can thus apply the theory from the early sections to propagate the uncertainty in the arguments of $ \mathcal F $ to the output using only few evaluations. As the output quantities are known to be positive as well, we shall assume log-normal distributions for these as well and fit them by computed means and variances.

\begin{figure}[!htb]
    \newcommand{\ww}{.425}
    \centering
    \includegraphics[width=\ww\textwidth]{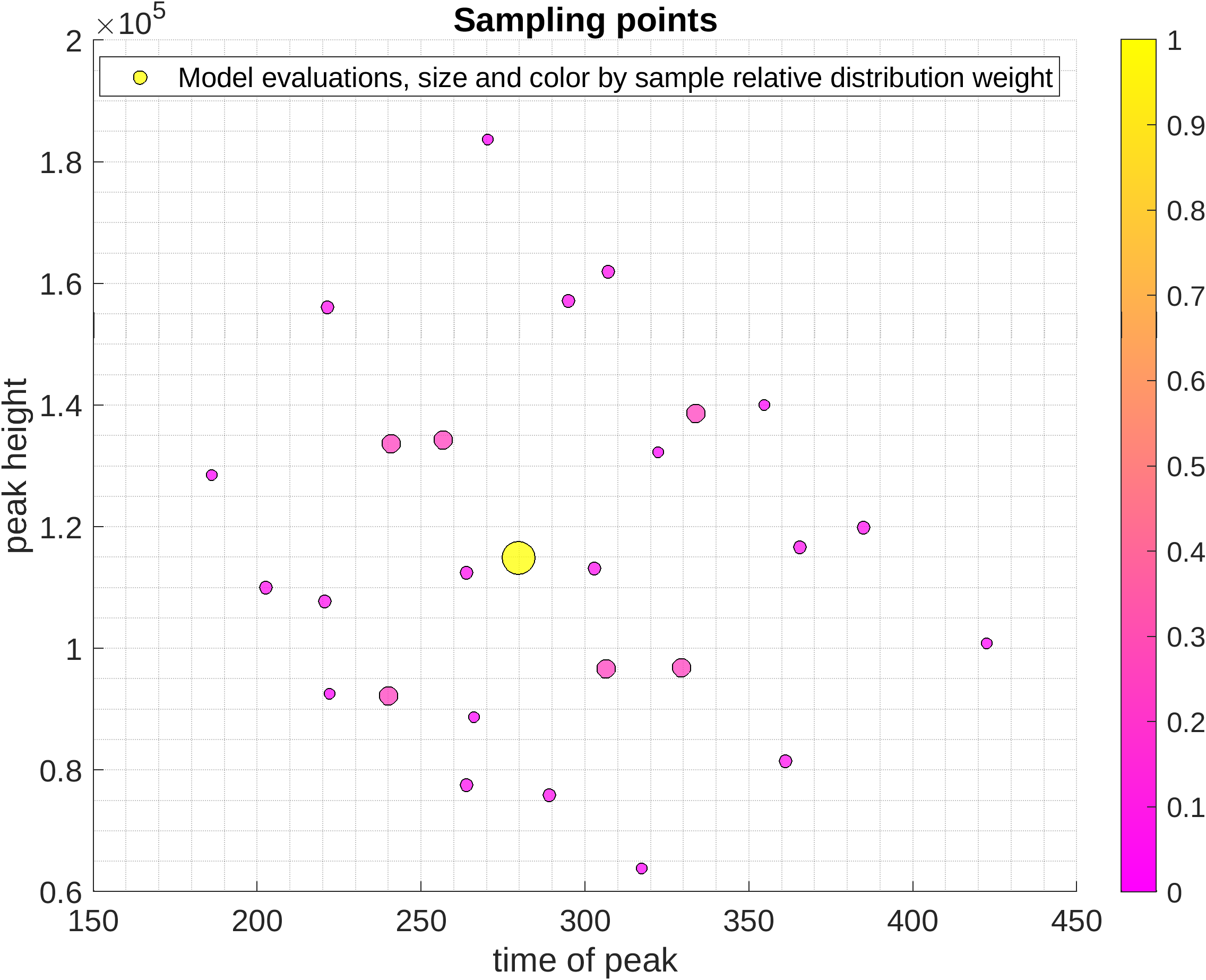}~%
    \includegraphics[width=\ww\textwidth]{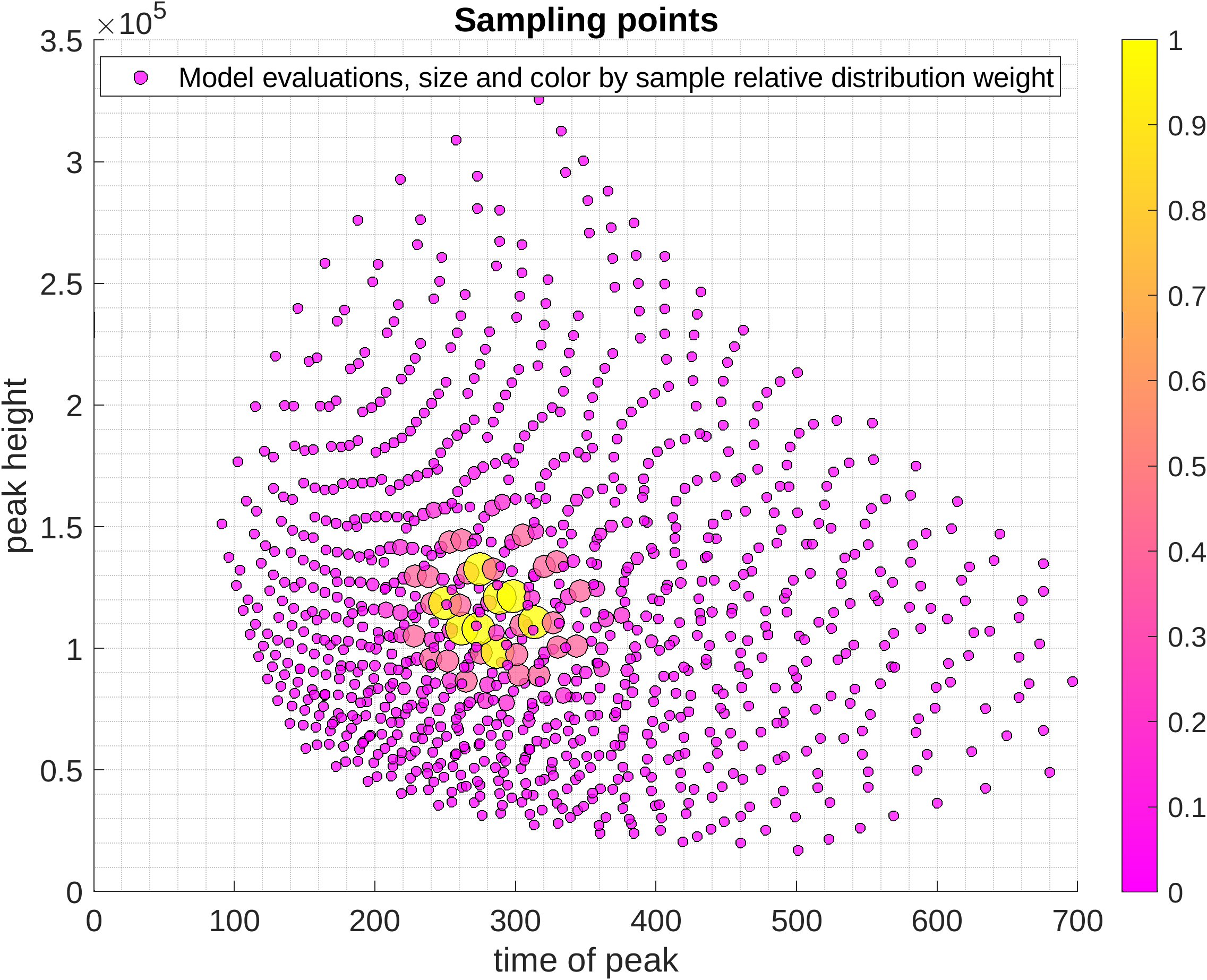}\\
    \includegraphics[width=\ww\textwidth]{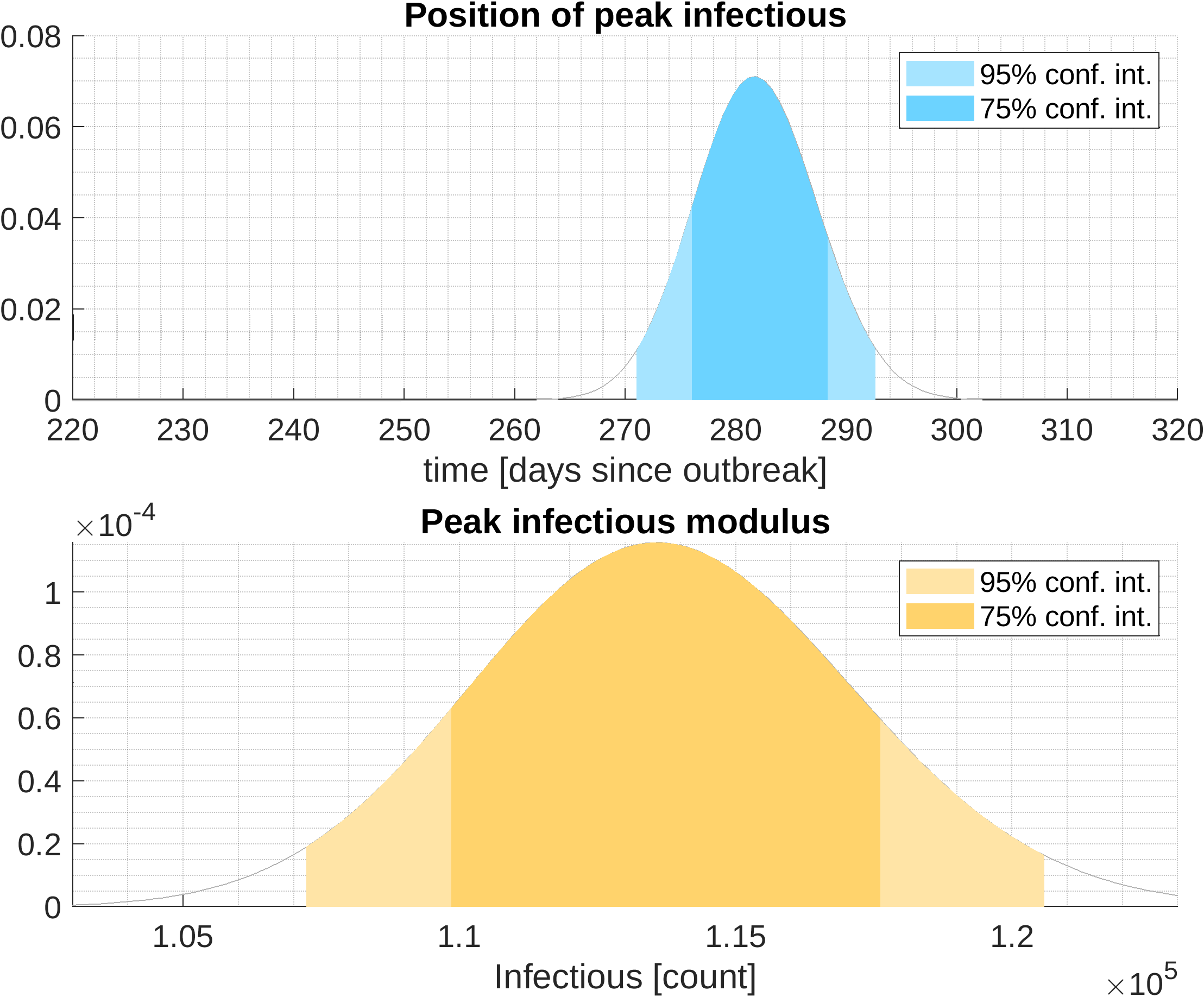}~%
    \includegraphics[width=\ww\textwidth]{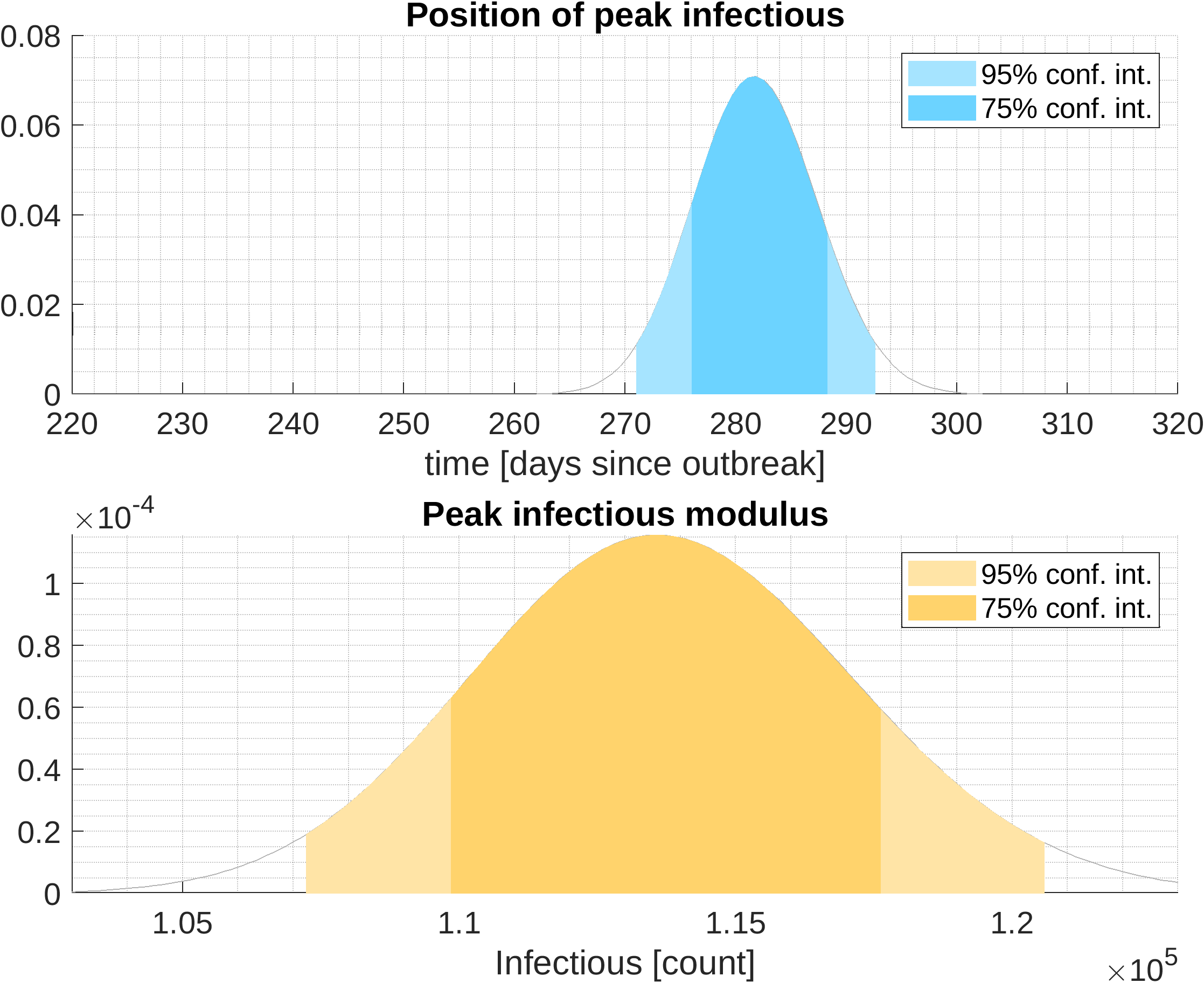}\\
    \includegraphics[width=\ww\textwidth]{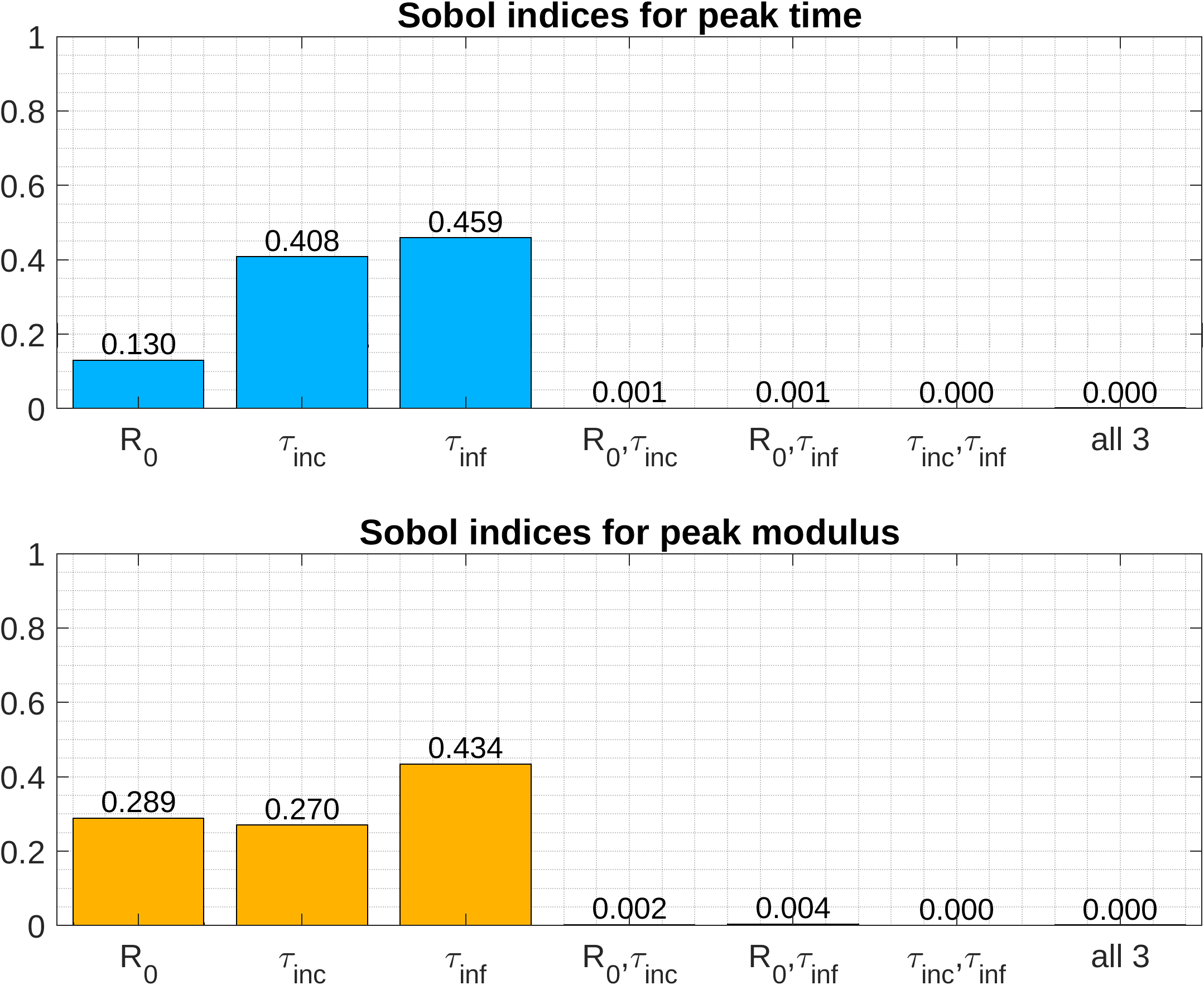}~%
    \includegraphics[width=\ww\textwidth]{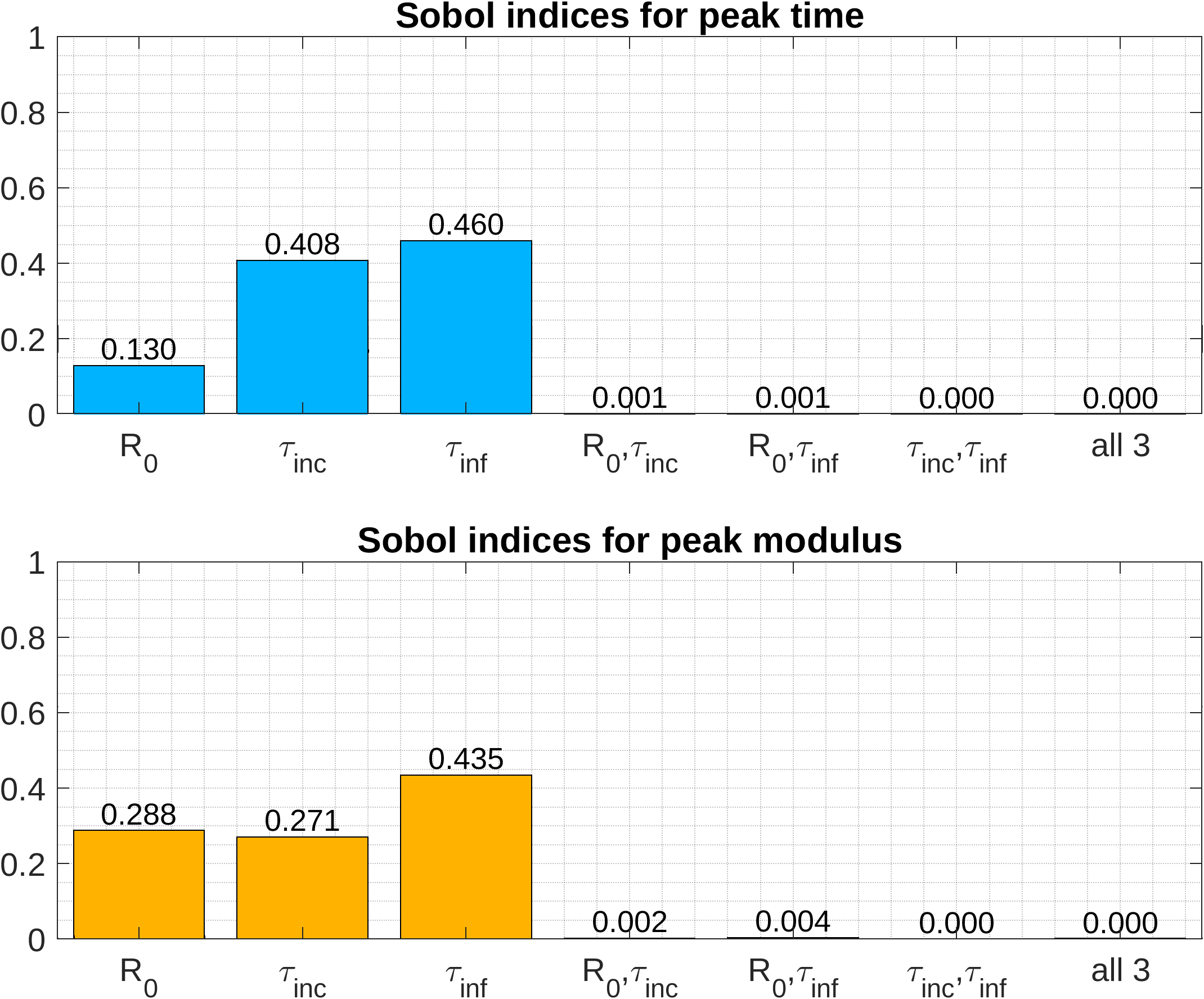}
    \caption{The peak of infectious individuals during a SEIR-model simulation. Each column corresponds to different applied quadrature degree. The top chart visualizes computed samples. The middle-charts the resulting distributions and the lower bar diagrams illustrate the variance contributions by the model parameters.}
    \label{fig:peak}
\end{figure}

The example can be seen in Figure \ref{fig:peak} where we have used the following hyper-parameters; chosen based on early reported numbers for Covid-19 with some level of contact restriction assumed.
\begin{center}
    \begin{tabular}{ c c c c }
        &  & mean & variance \\ 
        $ R_0 \sim$ & LogNormal & 1.4 & $0.025^2$ \\  
        $ \tau_{\text{inc}} \sim$ & LogNormal & 4.2 & $0.7^2$ \\
        $ \tau_{\text{inf}} \sim$ & LogNormal & 3.3 & $0.7^2$
    \end{tabular}
\end{center}
We compute the example with a low and a high number order of quadrature to illustrate how we may obtain quite accurate information with few model evaluations. In the left column of the figure we use 3rd order quadratures corresponding to $ 3^3 = 27 $ model evaluations, and in the right column we employ 10th order quadratures corresponding to $ 10^3 = 1000 $ model evaluations.

The top coordinate system of the figure illustrates all samples. The color and weight has been scaled by the probability of the corresponding values. The next axes shows the log-normal distribution for the peak time and the magnitude of the peak in infectious individuals. The last axes show the corresponding Sobol indices illustrating that with the selected values the variance of $ R_0 $ is not the primary concern if we wanted to narrow down the peak time further. 

We observe how the 27 model evaluations provides the same information as the 1000 model evaluations, showing that this problem is handled well already at this low number of evaluations.

\subsection{Case 2: Superspreaders}
Superspreaders are infected individuals who during an epidemic are responsible for the infection of a significantly larger amount of individuals than the observed average. Historical observations of diseases have shown that incidents with superspreaders play an important role\cite{lloyd2005superspreading}.

Various aspects play into causing an individual to become a superspreader, which may both be physiological and sociodynamic in nature. An individual exhaling an increased amount of pathogens relative to the norm could lead to a significantly larger number of infections during regular social interactions compared to a ``normal'' infectious individual. But it could also simply be the participation in a large scale social event for instance a party, concert or a festival, where the physical distancing may be very low and number of contacts proportionally higher, which results in mass infection. 

\begin{figure}[!hb]
\begin{center}
    \begin{tikzpicture}[scale=1,transform shape]
        \newcommand\s{1.75}
        \node[shape=circle,draw=black,minimum size=1cm,thick] (S) at (0,0) {$ S $};
        \node[shape=circle,draw=black,minimum size=1cm,thick] (E) at (\s,0) {$ E $};
        \node[shape=circle,draw=black,minimum size=1cm,thick] (I1) at (2*\s,0) {$ I_1 $};
        \node[shape=circle,draw=black,minimum size=1cm,thick] (I2) at (3*\s,0) {$ I_2 $};
        \node[shape=circle,draw=black,minimum size=1cm,thick] (W) at (4*\s,0) {$ W $};
        \node[shape=circle,draw=black,minimum size=1cm,thick] (H) at (4.66*\s,-0.66*\s) {$ H $};
        \node[shape=circle,draw=black,minimum size=1cm,thick] (C) at (6.33*\s,-0.66*\s) {$ C $};
        \node[shape=circle,draw=black,minimum size=1cm,thick] (R) at (7*\s,0) {$ R $};
        
        \path [->,thick, draw=black] (S) edge node[anchor=south,yshift=.3cm] {$\overline{\beta}(t)$} (E);
        \path [->,thick, draw=black] (E) edge node[anchor=south,yshift=.3cm] {$\sigma$} (I1);
        \path [->,thick, draw=black] (I1) edge node[anchor=south,yshift=.3cm] {$\gamma_1$} (I2);
        \path [->,thick, draw=black] (I2) edge node[anchor=south,yshift=.3cm] {$\gamma_2$} (W);
        \path [->,thick, draw=black] (W) edge node[anchor=south,yshift=.1cm] {$(1-z_1)\gamma_3$} (R);
        \path [->,thick, draw=black] (W) edge node[anchor=north east,yshift=.0cm] {$z_1\!\gamma_3$}  (H);
        \path [->,thick, draw=black] (H) edge node[anchor=south east,yshift=-.11cm,xshift=.1cm] {$(1-z_2)\alpha $} (R);
        \path [->,thick, draw=black] (H) edge node[anchor=north,yshift=.0cm] {$z_2\alpha$} (C);
        \path [->,thick, draw=black] (C) edge node[anchor=north west,yshift=.0cm] {$\zeta$} (R);

        \path [->,thick, draw=black, bend left=65, dashed] (I1) edge (.58*\s,-.15*\s);
        \path [->,thick, draw=black, bend left=65, dashed] (I2) edge (.42*\s,-.15*\s);
    \end{tikzpicture}
\end{center}
\caption{Illustration of expanded SIR-type model taking into account superspreaders; based on~\cite{sneppen2020impact}.}
\label{fig:sir-exp-model-ss}
\end{figure}
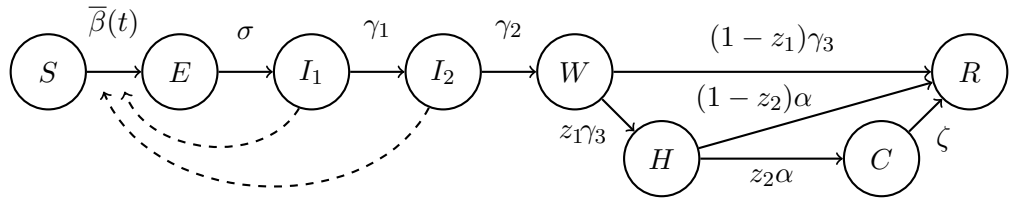

Inspired by \cite{sneppen2020impact} we attempt in this case to replicate some of their results in a computationally fast way using a SIR-type model. We employ the structure from their agent based model to construct the SIR-type model depicted in the diagram in Figure \ref{fig:sir-exp-model-ss}. In the diagram we have the following compartments: First, as in the former model susceptible and exposed. Then there are asymptomatic infectious $ I_1 $ and symptomatic infectious $ I_2 $. We note that this is a legacy structure from \cite{sneppen2020impact}, where it is used mostly for book keeping. Neither there nor here is behavior assumed to differ between the compartments. We have a (W)ait compartment, which signifies a short time, where the individual is either so sick that they have isolated themselves as to not infect anyone before admission to the hospital, or they are in not in non-infecting recovery. There is a branch with (H)ospitalized and (C)ritical care before all ending in the recovered/removed compartment.

The parameters choices are taken as in \cite{sneppen2020impact}, but we restate them for completeness in Table \ref{tbl:ss-model-params}.
\begin{table}[!ht]
\begin{center}
    \begin{tabular}{ c|c|c|c|c|c }
        $\sigma^{-1}$ & $\gamma_1^{-1}$ & $\gamma_2^{-1}$ & $\gamma_3^{-1}$ & $\alpha^{-1}$ & $\zeta^{-1}$ \\ \hline%
        1.2 & 1.2 & 3 & 2 & 5 & 12
    \end{tabular}
\end{center}
\caption{Superspreader model parameters~\cite{sneppen2020impact}. Units are in $ \left[\rm{days}\right]$.} \label{tbl:ss-model-params}
\end{table}
For $ z_1 $ and $ z_2 $ we compute them from the hospitalization rates listed in the \emph{Supplementary material Table~1} in \cite{sneppen2020impact} which comes from Norwegian data. We present the data in Table \ref{tbl:ss-model-hosp-prob} where $ d_i $, $ h_i $ and $ \kappa_i $ are the data rows. From these quantities $ z_1 $ and $ z_2 $ are computed as
\[
    z_1 = \sum_{i=1}^9 d_ih_i, \quad \text{and} \quad z_2 = \sum_{i=1}^9 \frac{d_ih_i}{z_1}\kappa_i.
\]
The expression for the last parameter $ \overline{\beta}(t) $ is given in \eqref{eq:ss-betaT} and \eqref{eq:ss-betaA}. The modelling approach is covered in Section \ref{sec:ss-beta-model}.
\begin{table}[!ht]
\begin{center}
    \begin{tabular}{ c|c|c|c|c|c|c|c|c|c }
        & 0-9 & 10-19 & 20-29 & 30-39 & 40-49 & 50-59 & 60-69 & 70-79 & 80+ \\ \hline%
        D ($ \{d_i\} $) [\%] & 10.9 & 11.9 & 13.3 & 11.7 & 13.6 & 13.6 & 11.7 & 8.9 & 4.4$^{\ast}$ \\%
        H ($ \{h_i\} $) [\%] & 0.001$^{\ast\ast}$ & 0.013 & 0.37 & 1.1 & 1.4 & 2.7 & 3.9 & 5.5 & 5.5 \\%
        C ($ \{\kappa_i\} $) [\%] & 5 & 5 & 5 & 5 & 6.3 & 12.2 & 27.4 & 43.2 & 70.9
    \end{tabular}\\[.25cm]
\end{center}
\begin{quote}
    \footnotesize\noindent $^\ast$) 0.1\% was added here since the numbers from the source table didn't actually add to 100\%.\\
    $^{\ast\ast}$) This number was 0 in the table, it is known that some kids end up hospitalized, so we changed it to a small but strictly positive value.
\end{quote}

\caption{Population distribution and hospitalization probability data~\cite{sneppen2020impact}. Legend: D: Distribution of the population; H: Probability of hospitalization; C: Probability of moving to critical care.} \label{tbl:ss-model-hosp-prob}
\end{table}

\subsubsection{Modelling varying infectivity} \label{sec:ss-beta-model}
We model a superspreaders by assuming a distribution of infectivity amongst individuals in the population. Consider the normalized population $ [0,1] $ and assign to each $ a \in [0,1] $ an infectivity $ \beta(a) $. That is, if $ U $ is some ordered set of possible infectivities and $ \psi $ is a probability measure on $ U $ with cumulative probability function $ \Psi $, then $ \beta(\cdot) = (1 - \Psi)^{-1}(\cdot) $. We assume that the population is ordered by decreasing infectivity; i.e. $ \beta(a) \leq \beta(a') $ for $ a < a' $. Assuming a well-mixed distribution such that infection is equally likely to hit any individual $ a \in [0,1] $ we may readily calculate the contribution to infection caused by the fraction $ p $ most infectious individuals, $ C_p $.

In a SIR-model the number of people getting infected at a time $ t $ is commonly, as seen in \eqref{eq:seir-ds}, of the form
\[
    \overline{\beta}\frac{I(t)S(t)}{N},
\]
where $ I(t) $ is the number of infected individuals, $ S(t) $ the number of susceptible individuals, and $ N $ is the population size. Here $ \overline{\beta} $ is the average infectivity; $ \overline{\beta} = \int_0^1\beta(a)\d a $.

The $ p $ most infectious individuals would then be contributing the fraction 
\begin{equation} \label{eq:ss-contrib}
    C_p = \overline{\beta}^{-1}\int_0^p \beta(a)\d a.
\end{equation}
This quantity informs the choice of probability measure $ \psi $ if one works under the scheme that superspreaders form some fraction $ p $ of the population and is responsible for infecting the fraction $ C_p $ of the population. Fixing these two quantities limits the admissible measures $ \psi $.

This way of modelling a variation in infectivity also admits fairly easy extensions to control scenarios where some rules may change behavioral dynamics over time. We may for instance consider a time-dependent infectivity
\[
    \overline{\beta}_{\text{restricted}}(t) = \int_0^1\phi(\beta(a),a,t)\d a,
\]
where $ \phi(b,a,t) $ is some restriction function describing a change in the behavior over time. In practice, however, $ \phi(b,a,t)\equiv \phi(b,t) $ will typically be independent of $ a $ as we cannot feasibly identify an individual as more infectious than another until after the fact. And so we have to make rules that are uniform for everyone. A simple example could be a strict limitation in how many individuals anyone meet, which could be crudely modelled as 
\begin{equation} \label{eq:restriction}
    \phi(b,a,t) = \min(b,c(t)),
\end{equation}
where $ c\colon \R_+ \to \R_+ $ is some time-dependent upper bound.

Of course, if $ \phi(b,a,t) \equiv \phi(a,t) $ is independent of $ b $, we may impose any kind of other ordering on the population, e.g. by age, and apply hard restrictions based on that one. But then we lose all information from the infectivity $ \beta $, which might be undesirable.

We take for our model $ \beta(a) $ as a simple piecewise constant function
\begin{equation} \label{eq:ss-betaA}
    \beta(a) = \begin{cases}
        sA & \text{if $ a < p $}, \\
        s & \text{if $ a \geq p $},
    \end{cases} \quad a \in [0,1].
\end{equation}
Here $ p $ is the assumed concentration of superspreaders; e.g. if we assume 10\% are superspreaders $ p = 0.1 $. $ sA $ is the infection rate for superspreaders and $ s $ the infection rate for the remaining population. It is an easy calculation that $ C_p $ from \eqref{eq:ss-contrib} is independent of $ s $, so choosing an assumed $ (p,C_p) $ pair determines $ A $ and choosing $ s $ then determines the mean rate $ \overline{\beta} $.

We shall model a social restriction as a hard cap on the amount of individuals any single person gets to interact with. We model this with a restriction function as in \eqref{eq:restriction}; i.e.
\begin{equation}
    \overline{\beta}(t) = \int_0^1 \min(\beta(a),c(t))\d a. \label{eq:ss-betaT}
\end{equation}
with $ c $ being a piecewise constant function which changes values at approximately 1) the time of the Danish lockdown, 2) the timing of the Danish reopening's phase 1 (about a month later), 3) the timing of the Danish reopening's phase 2 (another about 40 days later).

\subsubsection{Fitting the model}
From an assumption about the prevalence of superspreaders we first fit $ \beta $'s scaling parameter $ s $ and an initial condition $ I_0 $ for the epidemic from hospital admission by day\footnote{Danish data available from SSI (www.ssi.dk), the Danish Ministry of Health. The data was public and accessed on June 14th, 2020; it does not remain available anymore. The used data file is available as a CSV-file with the codes in the GitLab repository.} (only for the pre-lockdown part of the data set) and an assumption of an unmitigated growth rate at about 23\% per day\cite{sneppen2020impact}. For the initial condition we fit a number $ I_0 $ and we assume that $ S(0) = N - I_0 $, $ E(0) = \frac{I_0}{2} $, $ I_1(0) = \frac{I_0}{3} $ and $ I_2(0) = \frac{I_0}{6} $. The remaining compartments start at 0. The problem is formulated as
\begin{equation}
    \argmin_{s,I_0} \frac{w_{0}}{2}|\mathcal G(s,I_0) - 0.23|^2 + \alpha\frac{w_{1}}{2}\|\mathcal H_{t<t_1}(s,I_0)-H_{\text{ssi},{t<t_1}}\|^2,
\end{equation}
where the weights $ w_{0}^{-1} = 0.23^2\;\left[\frac{\rm{persons}^2}{\rm{day}^2}\right]$ and $ w_{1}^{-1} = \|H_{\text{ssi}}\|^2\;\left[\rm{persons}^2\right] $ balance the widely different scales of the two terms, and $ H_\text{ssi} $ is the data set of newly admitted hospitalized by day. $ \mathcal G $ computes the average initial daily growth rate and $ \mathcal H $ computes the newly admitted hospitalizations from the model. By the subscript ${t<t_1}$ we mean only the part of the data corresponding to this constraint; $ t_1=16 $ before which $ \mathcal H $ is independent of our restriction function. We chose $ \alpha = 0.01 $.

Using the now determined quantities $(s,I_0)$ we fit the three restriction levels in $ c(t) $ from the whole data set of hospital admission by day. Fixing $ (t_1,t_2,t_3) = (16,46,86) $ we have
\begin{align}
    c(t) = \begin{cases}
        1 & \text{if $ t \leq t_1 $}, \\
        c_1 & \text{if $ t_1 < t \leq t_2 $}, \\
        c_2 & \text{if $ t_2 < t \leq t_3 $}, \\
        c_3 & \text{if $ t_3 < t $}.
    \end{cases}
\end{align}
Then the parameter fitting problem becomes
\begin{equation}
    \argmin_{c_1,c_2,c_3}\frac{1}{2}\|\mathcal H(c_1,c_2,c_3)-H_{\text{ssi}}\|^2 - w_2(\min(0,c_2-c_1)+\min(0,c_3-c_2)),
\end{equation}
where $ w_2 $ is some arbitrary large number so the last term forms a soft constraint enforcing $ c_1 < c_2 < c_3 $.

\begin{figure}[htb]
    \centering
    \includegraphics[width=.5\textwidth]{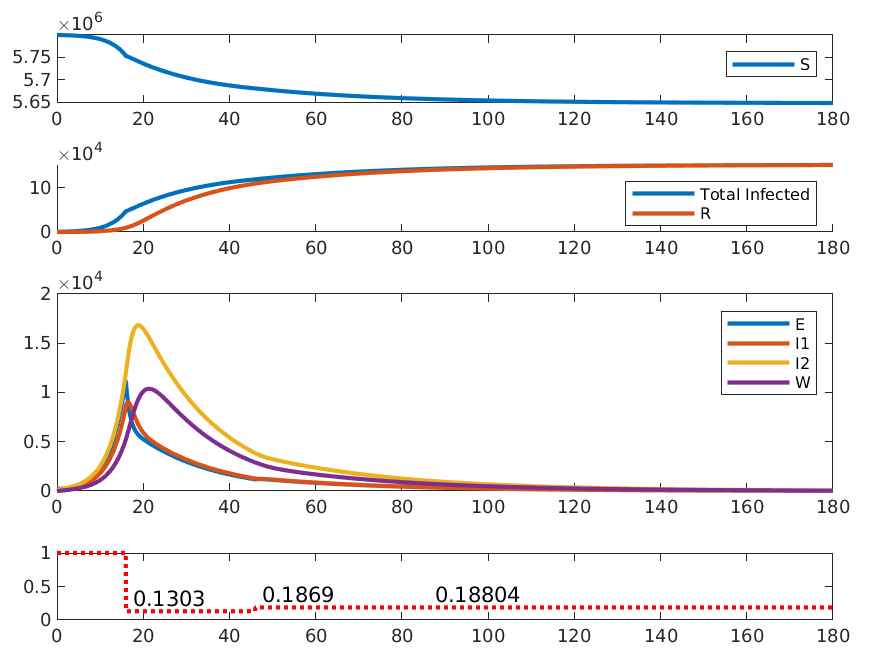}~%
    \includegraphics[width=.5\textwidth]{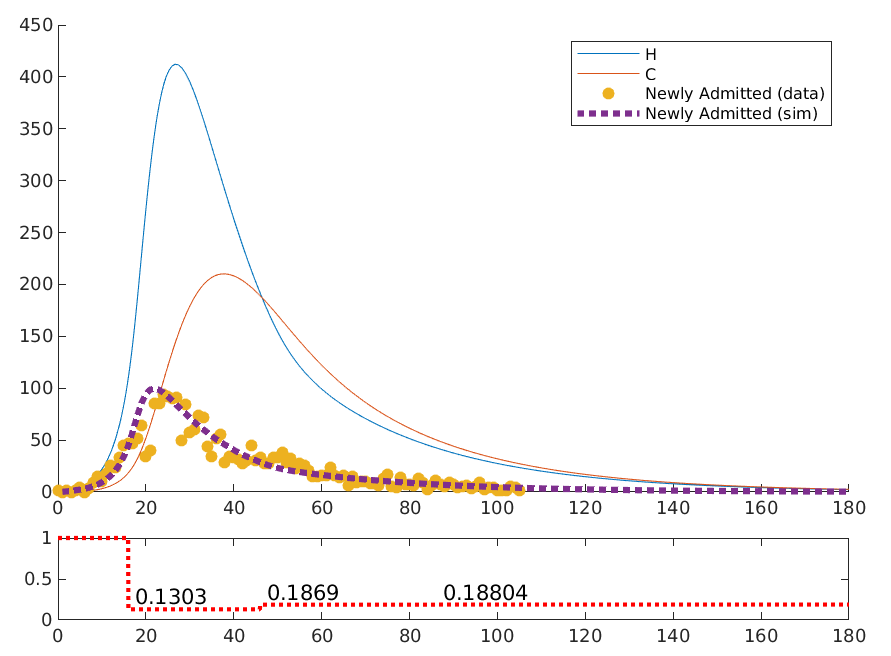}
    \caption{Superspreader case simulation using fitted data. The time-evolution of different compartments have been visualized grouped by their relative $y$-scale. $x$-scale units are in days and $ y $-scale units are counts. The dotted red curve seen lowest in both charts is the active restriction. The yellow dots in the right chart is newly hospital admissions by day from the Danish authorities.}
    \label{fig:ss-model-fit01-sim}
\end{figure}

Assuming $ (p,C_p) = \left(\frac{1}{10},\frac{4}{5}\right) $, i.e. that only 10\% contribute 80\% of all infections, the above fitting schemes resulted in 
\[
    s = 0.602\;\left[\frac{1}{\rm{day}}\right], \quad I_0 = 473.572\;\left[\rm{persons}\right], \quad \text{and} \quad (c_1,c_2,c_3) = (0.130, 0.187, 0.188).
\]
These results depended slightly on the choice of initial condition but the differences were on the order of $ 10^{-3} $. The simulation, when done using these data, may be viewed in Figure \ref{fig:ss-model-fit01-sim}.

We note that the difference between restriction levels $ c_2 $ and $ c_3 $ is almost insignificant. There are likely various reasons for this. In the model in \cite{sneppen2020impact} they have different society structures which they can close down. Comparatively, we only really have one here. A possible explanation might be that the phase 2 reopening didn't really affect the overall amount of contacts for people. Of course, this could also be a data deficiency as small variations of $ c_3 $ has proven to not change the optimization functional significantly.

\subsubsection{Adding uncertainty}
\begin{figure}[htb]
    \centering
    \includegraphics[width=.80\textwidth]{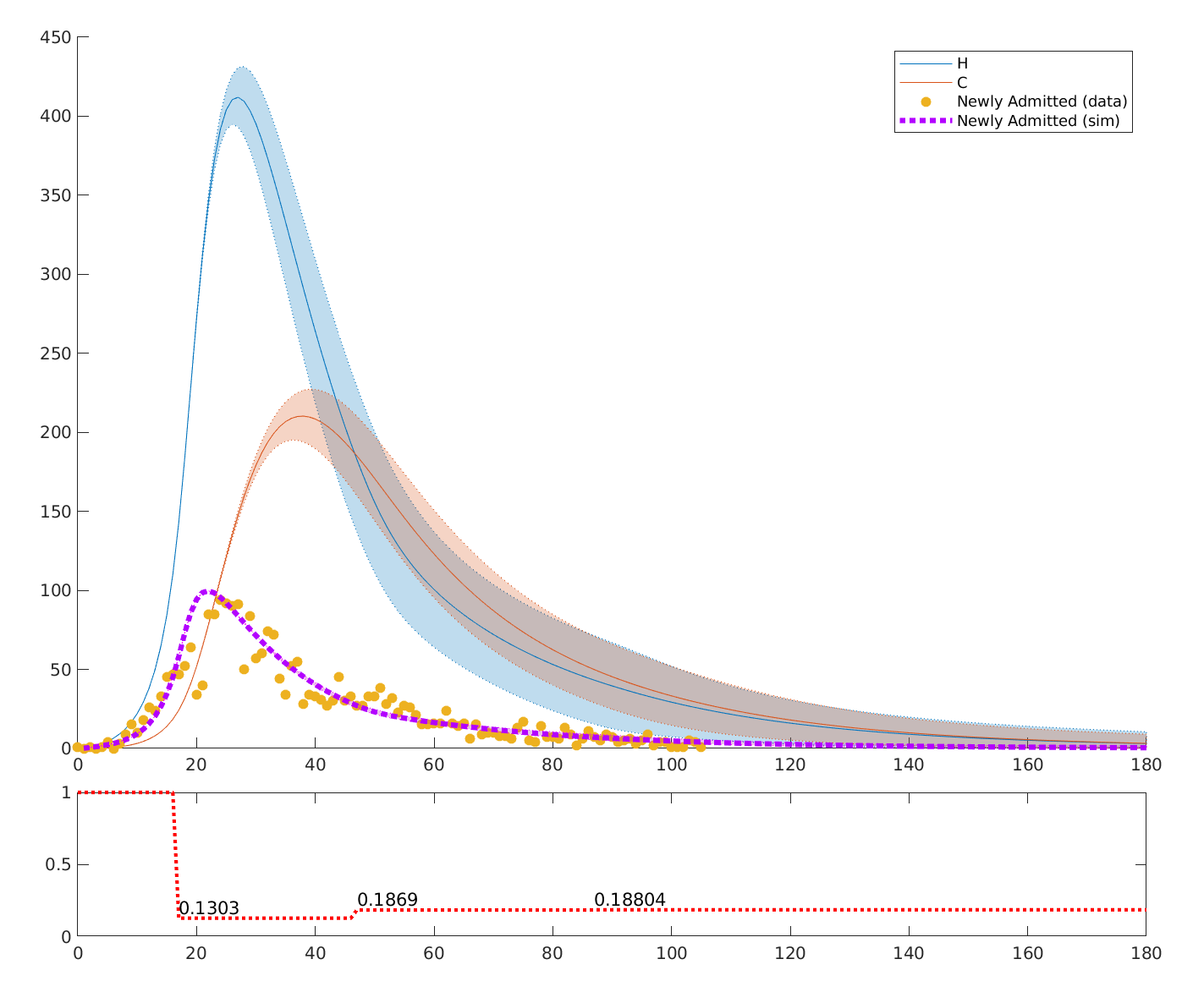}
    \caption{Superspreader case simulation using fitted data taking into account uncertainty. Hospitalized (H), critical care (C) and newly hospitalized are shown with confidence intervals. $x$-scale units are in days and $ y $-scale units are counts. The dotted red curve is the active restriction. The yellow dots are newly hospital admissions by day from the Danish authorities.}
    \label{fig:ss-model-fit01-conf}
\end{figure}

Assuming a level of uncertainty in the fitted restriction levels we may compute confidence intervals for the model. In Figure \ref{fig:ss-model-fit01-conf} we assume normally distributed priors for the restriction levels with means $ c_i $, $ i=1,2,3 $, and relatively scaled standard deviations of $ 0.1c_i $, $ i=1,2,3 $. Assuming the posteriors may be approximated reasonably by a truncated normal distributions, 95\% confidence intervals are visualized. We see that with uncertainty of this level on the parameters the development is expected to keep declining.

\begin{figure}[htb]
    \centering
    \includegraphics[width=.60\textwidth]{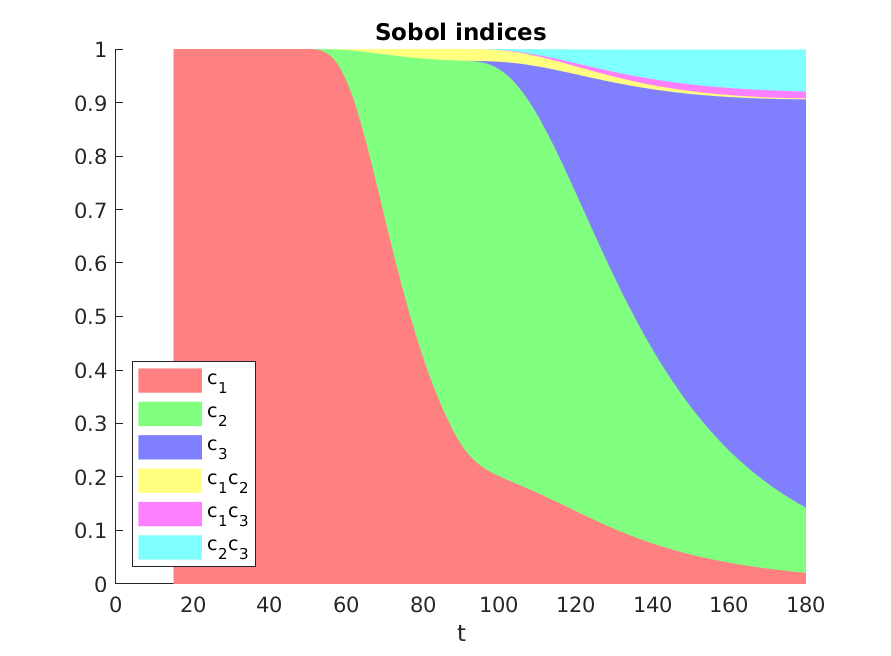}
    \caption{The Sobol indices evolution over time in the superspreader case simulutation using fitted data with added uncertainty. The $x$-axis is in units of days. At each time the color distribution above determine the part of the variance contributed by each parameter; the corresponding parameter listed in the legend-box. The $c_ic_j$ parts are the joint variance contributions of $ c_i $ and $ c_j $, as is visible, $ c_1 $, $ c_2 $ and $ c_3 $ are mostly unrelated here.}
    \label{fig:ss-model-fit01-sobol}
\end{figure}

The evolution of the Sobol indices over time is drawn up in Figure \ref{fig:ss-model-fit01-sobol} illustrating the variance contributions from the parameters, which show as expected how $ c_1 $ is the most important initially but is gradually taken over by $ c_2 $ and then $ c_3 $ in the later stages. Notably $ c_1 $ remains fairly important even during the time span where $ c_2 $ controls the level of interaction, and likewise $ c_2 $ into the time span where $ c_3 $ is active.

\section{Conclusion}
We have in this summarized the key elements of generalized Polynomial Chaos and the related Polynomial Chaos Expansions, as well as their applications to the efficient quantification of uncertainty in models; both basic statistical properties and the Sobol indices. The novelty of this work lies in the application of these techniques to epidemic models; here applied to official Danish data from the Covid-19 epidemic, which struck in 2020 and remains an issue still in 2021. We find that taking uncertainty into account in predictions of these types are of tremendous value and utmost importance, and that these tools are well suited in the field of epidemic modelling. We thus recommend using these tools, which have so far remained outside the field, for their efficiency. Not to replace existing tools, but to provide a wider array of options suitable for different purposes.

\section*{Acknowledgements}
This work was supported by the project \emph{Estimation, Simulation, and Control for Optimal Containment of COVID 19} from the Novo Nordisk Fonden; project no. NNF20SA0063089 (Application no. 67). BCSJ was supported by the Academy of Finland (grant no. 320022).

\bibliographystyle{plain}
\bibliography{bibliography}